\documentclass[10pt,letterpaper]{article}

\usepackage{cogsci}

\cogscifinalcopy 
\usepackage{svg}
\usepackage[
    backend=biber,
    style=apa,
    natbib=true,
    doi=false,
    isbn=false,
    url=false,
]{biblatex}
\usepackage{pslatex}
\usepackage{float} 

\usepackage{xcolor}
\usepackage{ragged2e}
\usepackage{blindtext}
\usepackage{hyperref}
\usepackage{graphicx}

\title{Framing Perception: Exploring Camera Induced Objectification in Cinema}
 
\author{{\large \bf Parth Maradia (parth.m@research.iiit.ac.in)} \\
    Cognitive Science Lab, IIIT-H, Telangana, India, 500032
  \AND {\large \bf Ayushi Agarwal (ayushi.a@research.iiit.ac.in)} \\
    Cognitive Science Lab, IIIT-H, Telangana, India, 500032
    \AND {\large \bf Srija Bhupathiraju (srija.bhupathiraju@research.iiit.ac.in)} \\
    Cognitive Science Lab, IIIT-H, Telangana, India, 500032
    \AND {\large \bf Kavita Vemuri (kvemuri@research.iiit.ac.in)} \\
    Cognitive Science Lab, IIIT-H, Telangana, India, 500032 }

\begin{document}

\maketitle

\begin{abstract}
This study investigates how cinematographic techniques influence viewer perception and contribute to the objectification of women, utilizing eye-tracking data from 91 participants. They watched a sexualized music video (SV) known for objectifying portrayals and a non-sexualized music video (TV). Using dynamic Areas of Interests (AOIs)—head, torso, and lower body—gaze metrics such as fixation duration, visit count, and scan paths were recorded to assess visual attention patterns. Participants were grouped according to their average fixations on sexualized AOIs. Statistical analyses revealed significant differences in gaze behavior between the videos and among the groups, with increased attention to sexualized AOIs in SV. Additionally, data-driven group differences in fixations identified specific segments with heightened objectification that are further analyzed using scan path visualization techniques. These findings provide strong empirical evidence of camera-driven gaze objectification, demonstrating how cinematic framing implicitly shapes objectifying gaze patterns, highlighting the critical need for mindful media representation.

\end{abstract}

\section{Introduction}

Across global media landscapes, women are frequently portrayed to emphasize their physical attributes, perpetuating sexual objectification—reducing them to sexual body parts devoid of agency \citep{FredricksonRoberts1997, Bartky1990}. Cinema, as a powerful medium of storytelling, shapes societal attitudes and perceptions through camera angles, editing, and narratives \citep{Seeley2013CognitivismPA, Singh2024THEIO}. By emotionally and cognitively engaging viewers, films and music videos can induce or reinforce social norms that direct their gaze and behavior \citep{Carmi2006VisualCV, Smith2010AttentionalSI, Karsay2018SexualizingMU, Karsay2020SexuallyOP}. A key concern is the objectified portrayal of women through revealing attire, extreme skin exposure, suggestive choreography, and focus on sexualized body parts, which commodify their bodies \citep{Flynn2016ObjectificationIP}. Repeated exposure to such imagery can cause people to internalize such seemingly harmless perspectives, which fosters self-objectification, negative body image, and interpersonal objectification, potentially leading to harmful behaviors \citep{Aubrey2011SexualOI, Ward03052016}.

Cinematic theory posits that the “male gaze” \citep{Mulvey1975VisualPA} fragments female body for voyeuristic and fetishistic pleasure. This objectification is reinforced by the cinematic fourth wall. In Indian cinema, this is particularly evident in “item songs” \citep{Slatewala2019ObjectificationOW, mathew2023has} which originated in the 1970s to eroticize women's bodies for male pleasure \citep{kumari2022objectification}. These songs, often force fitted into movies to attract viewers, feature provocative choreography, scanty attire, and sexualized camera angles, reducing women to objects of visual consumption \citep{brara2010item, dwivedi2017sexual, jain2019sexuality, kamble2022portrayal}. The colloquially used term “item” dehumanizes women, labeling performers dancing in sexually suggestive ways as “item girls” meant to gratify male desire. Despite their popularity in social settings, such songs face little criticism for their damaging impact \citep{lamb2019sexualization, sharma2021content}. Studies show 60\% of music videos focus on sexual themes \citep{gruber2000adolescent}, and 89.6\% of item songs since 2010s feature provocatively dressed performers \citep{Slatewala2019ObjectificationOW}. While objectification in Western media is widely studied, its role in South-Asian cinema, particularly Indian, remains underexplored \citep{loughnan2010objectification, bhandari2018commodification,  zoon2019gender, nath2021review, aksar2022cinematography}.

Previous research relying on static images or self-reports fails to fully capture subconscious biases and attentional shifts elicited by dynamic stimuli like films \citep{negi2020fixation, Ward03052016, vaes2011sexualized}. Eye tracking technology offers real-time nuanced insights into gaze behavior revealing how visual attention is directed toward sexual stimuli \citep{wenzlaff2016video, Karsay2018AdoptObj, gervais2013my, duchowski2017eye, bhupathiraju2024objectifying}. Metrics like fixation duration, visit count, and scan paths quantify objectification, uncovering cognitive and subconscious processes guiding attention. Our study employs eye-tracking to differentiate between camera-guided objectification driven by cinematic techniques, and non-camera-guided objectification driven by viewer's internal biases. This distinction is critical for understanding how films uniquely influence perceptions of women. Our data-driven segmentation approach, inspired by \citet{onwuegbusi2022data}, isolates moments when cultural and cinematic cues intensify focus on sexualized AOIs, overriding individual differences, capturing temporal nuances and saccadic frequencies overlooked by static stimuli or self-reported measures. Scan paths provide a dynamic view of sequential gaze patterns, offering comprehensive insights into how camera techniques shape perceptions \citep{jacob2003eye, takeuchi2012scan}.

\section{Scope}

This study expands objectification research into the realm of dynamically edited, culturally specific media, focusing on Indian music videos “item songs” known for their provocative choreography, revealing attire, and strategic camera angles that sexualize female performers \citep{kapoor2019self, Slatewala2019ObjectificationOW}. We examine how contextual cinematic framing and editing influence cognitive and perceptual biases \citep{Aubrey2011SexualOI, Karsay2018AdoptObj} by analyzing a sexualized music video (SV) and a non-sexualized traditional video (TV). While the SV condition employs sexualized framing to emphasize female body parts, the TV condition is distinguished by modest attire, graceful movements, focusing on cultural expression rather than sexualization. Using eye-tracking, we analyze fixation duration, visit count, and scan paths to explore the effect of sexualized framing on gaze behavior  \citep{Karsay2018AdoptObj}. Our data-driven segmentation identifies intervals where objectifying gaze patterns intensify across subjects. While auditory and bodily movement may also influence gaze, we focus on the primary dancers, excluding background performers. Conducted within an Asian-Indian demographic, this research extends objectification studies to a non-Western context and lays the foundation for more nuanced future investigations into how cinematic techniques shape cognition and perception over time.

Building on objectification theory \citep{FredricksonRoberts1997, Bartky1990} and the documented influence of camera framing in directing objectifying attention \citep{Aubrey2011SexualOI, Ward03052016, Karsay2018SexualizingMU}, we formulate that SV induces a measurable intensification of objectifying gaze patterns in participants resulting in \textbf{RQ1:} Does camera framing in SV result in increased fixation duration and visit counts on sexualized AOIs (torso, lower body) compared to TV? Drawing from research on shot composition and affective impacts on visual processing \citep{Canini2011AffectiveAO}, we explore \textbf{RQ2}: Are there differences in the viewer’s gaze behavior while viewing close-up vs. long-shot composition? We anticipate that close-up shots, which restrict the visual field and emphasize specific details, may enhance attentional synchrony across subjects \citep{Smith2010AttentionalSI} and potentially increase the salience of sexualized AOIs, thereby promoting greater gaze synchronization among viewers. Rapid editing techniques and targeted camera framing in SV create a highly constrained visual environment, directing participant’s attention to specific body parts. Building on this we hypothesize, \textbf{H1:} SV will elicit greater gaze synchronization towards sexualized AOIs compared to TV, indicating stronger exogenous attentional control driven by cinematic techniques.

\section{Methodology}

\subsection{Participants}

Ninety-one self-identified heterosexual engineering students (M=68, F=23; ages 18-25, mean age = 21.6) participated in the study approved by the Institute’s Ethics Committee. Informed consent was obtained, ensuring participant’s confidentiality and their right to withdraw. One participant's data with a low ($<$40\%) eye-tracking sampling rate was excluded.


\subsection{Apparatus}

Eye movements were recorded using a Tobii X-120 eye tracker at 120 Hz. Stimuli were displayed on a 22-inch LCD monitor (1920×1080 px, 60 Hz refresh rate), with participants seated 51 to 71 cm from the display for optimal data quality. The setup minimized head movement to ensure calibration accuracy (nine-point), and gaze data was captured using Tobii Studio Pro software (v 3.3.2).

\subsection{Stimuli}

\subsubsection{Music Videos:} To identify stimuli for our study, we conducted another online pilot study with 84 participants (F=30, mean age=31.43; M=54, mean age=36.53; 18-67 years) who rated three sexualized music videos from commercial cinema on a 5-point Likert scale based on their perceived arousal. The video with the highest rating (3.30) of perceived arousal was selected as SV—\href{https://www.youtube.com/watch?v=5nPNv6_d_kI}{Aashiq Banaya Aapne}, an item song featuring provocative dancing and revealing attire. Another video—\href{https://www.youtube.com/watch?v=tzRFLMn4kpM}{Pinga} from the movie “Bajirao Mastani” was selected as TV, featuring traditional Indian dance style “Lavani”. Both videos featured only female characters to control for gender effects. Additionally, female participants completed the Interpersonal Sexual Objectification Scale (ISOS) and the Objectified Body Consciousness Scale (OBCS), with mean scores of 33.50 and 20.43 respectively while male participants completed a modified ISOS-Perpetrator (ISOS-P) scale, with a mean score of 69.87 \citep{gervais2018development, mckinley1996objectified, kozee2007development}. These scores suggest moderate levels of experienced and internalized sexual objectification among women, and a relatively high tendency to objectify among men. 



\subsubsection{Images:} Twelve grayscale images of female models dressed in four types of attire were selected from online shopping websites. The attires were categorized into two groups: (1) Traditional—Saree and Salwar, and (2) Non-Traditional/Western—Shirt-Pant and Short Dress (above knee length). Three images for each of the 4 attires were included, resulting in a total of twelve images. Only images of thin-bodied, as defined by \cite{Karsay2018AdoptObj}, female models were included as prior research indicates that such females conform to media-promoted sexualized notions of women and are more vulnerable to media objectification \citep{gervais2013my, kapoor2019self, gervais2011you}. The priming effect analysis is not included in this paper.

\subsection{Procedure}

This study employed a within-subject design \citep{Karsay2018AdoptObj, bhupathiraju2023objectifying}, to investigate camera-induced objectifying gaze behavior. Participants were randomly assigned to both video conditions (TV and SV) in a counterbalanced order to account for within-subject variability and mitigate potential order effects. As shown in Figure 1, to mask the true objective of observing natural video viewing behavior, each video presentation was followed by an attractiveness rating task of models in different attires. Only gaze data from the videos were analyzed for the current study.

\begin{figure}[htbp]
    \centering
    \includegraphics[width=0.99\columnwidth]{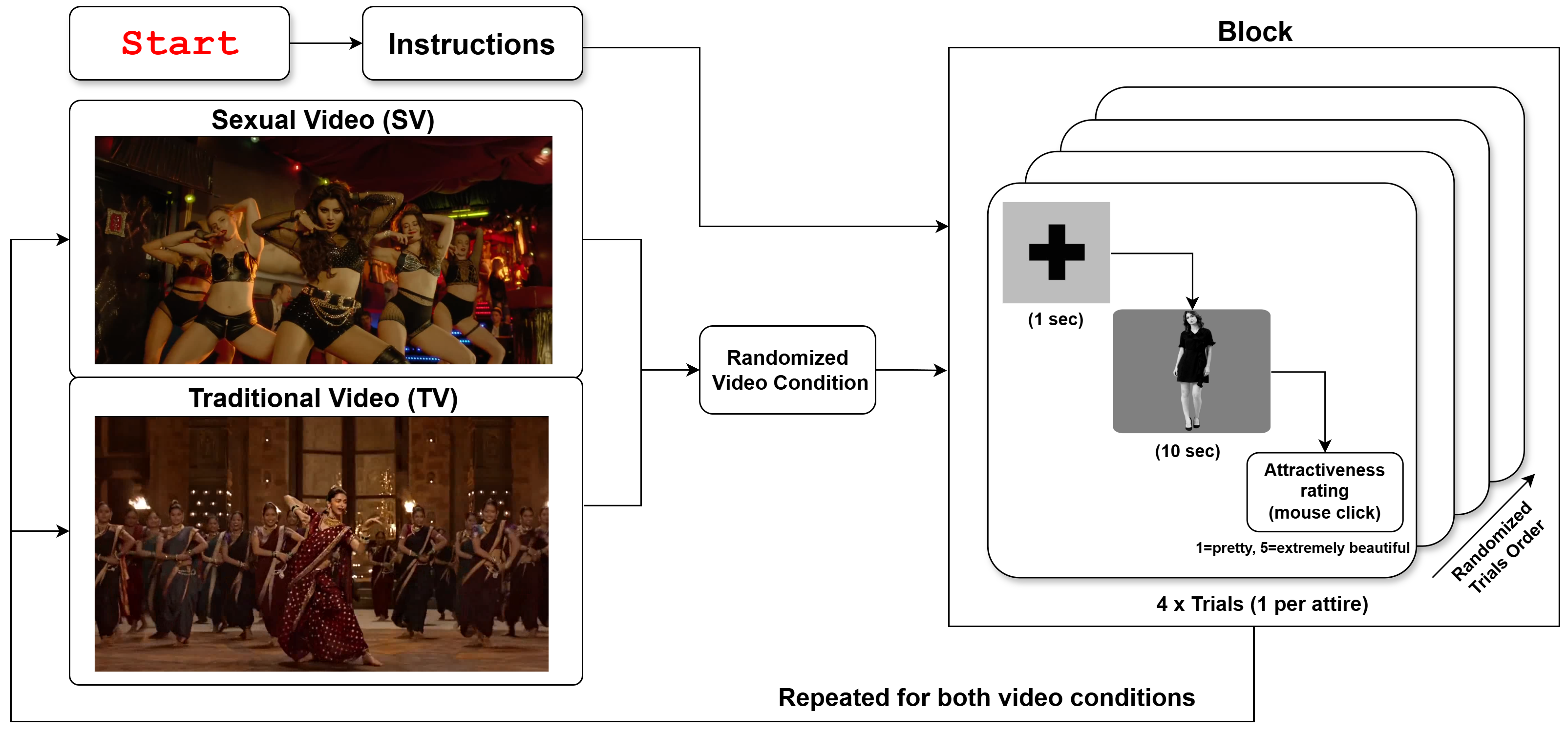}
    \caption{\textbf{Experimental paradigm.}}
    \label{fig:exp_pdgm}
\end{figure}

\subsection{Data Analysis}

\subsubsection{Areas of Interest (AOIs) Definition:}
Prior to our two-stage data analysis (Figure 2a), we selected three AOIs for our study, Head, Torso/chest, and Lower Body(hips/legs). While the head is not typically sexualized, it is prominently viewed during objectification, particularly in Indian/South-Asian demographics \citep{bhupathiraju2024objectifying}. We categorized the Torso and Lower Body as Sexualized AOIs and manually marked all three AOIs for the images and music videos (Figure 2b). The Dynamic AOI Tool from Tobii Studio Pro was used to mark AOIs in keyframes of the video, which are further extrapolated to ensure all frames have accurate AOI markings.

\subsubsection{Gaze Metrics:}
Normalized gaze metrics such as fixation duration, visit count, and scanpaths were extracted and analyzed (refer to Figure 2a) to capture the depth, frequency, and order of gaze points to investigate the visual engagement with the dynamic stimuli. While fixation duration measures sustained attention on specific AOIs reflecting the depth of cognitive engagement, visit count captures the recurrence of gaze highlighting the elements that repeatedly draw visual attention \citep{kimeyetrack2012, negi2020fixation}. However, due to strongly asymmetrical patterns of viewing—where certain AOIs under specific camera techniques garner substantially higher attention—our data distributions deviate markedly from normality (confirmed by Shapiro-Wilk tests, p $<$ 0.05). This complexity necessitates the use of non-parametric statistical methods to ensure robust comparisons across conditions, accommodating the skewed nature of the observed gaze behavior.
\begin{figure}[h]
    \centering
    \includegraphics[width=0.98\columnwidth]{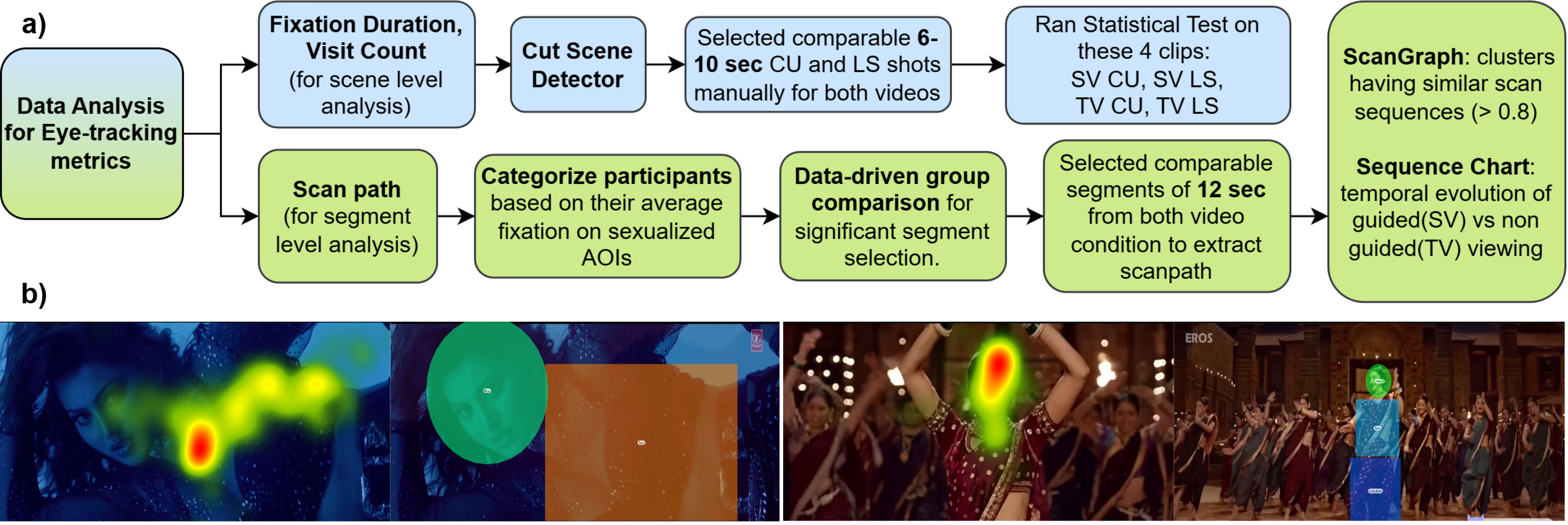}
    \caption{a) Flowchart of the two-stage analysis of gaze metrics. b) Sample frames from SV (left) and TV (right) with defined AOIs and heatmap overlays showing participants' aggregated gaze distribution.}
    \label{fig:data_analysis_plot}
\end{figure}

\subsubsection{Selection of Scenes and Segments from Videos:}
We identified and extracted comparable Close-up (CU) and Long-shot (LS) scenes with durations between 6-10 seconds from both the videos. This selection allowed us to focus on specific camera techniques that could influence gaze behavior related to objectification. In CU scenes, only the head and torso were visible, so we excluded the lower body AOI only from further CU scene analysis.

We initially divided participants into two equal groups (high and low objectifying) ranking them based on their average fixation counts on sexualized AOIs in the SV condition. However,  based on the sequence chart analysis for manually selected random segments (from both SV and TV) of comparable length, this binary grouping failed to capture nuanced variations in the low objectifying group, where some exhibited aversive gaze despite the camera direction in SV. Therefore, we refined our approach by dividing participants into three equal groups (high, medium, and low objectifying). This categorization, validated through sequence chart analysis and a data-driven segmentation approach inspired by \cite{onwuegbusi2022data}, allowed us to identify significant video segments eliciting distinct gaze patterns. We conducted independent t-tests on the Euclidean distances of individual gaze from their respective group centers, applying Bonferroni correction. Our segmentation approach not only quantifies within-group variations in gaze position but also distinctly marks frames showing significant inter-group differences, offering empirical support for identifying moments of strong or divergent reactions to specific scenes.

\textbf{ScanGraph:}
Post identification of significant video segments of 12 seconds each, \citet{Dolezalov2016ScanGraphAN}’s ScanGraph was employed to analyze differences in gaze sequences within and between groups. We constructed groupwise scanpath strings by assigning each AOI a unique letter (Head = A, Torso = B, Lower Body = C, No AOI Fixation = X), primarily to ensure that each scanpath string were of the same length and captured gaze directed towards and away from the central subject. Using Levenshtein string-edit distance with a similarity threshold of 0.80, participants with similar scanpath sequences were visualized as cliques in the ScanGraph. This analysis allows us to uncover underlying patterns in scanning behavior among different clusters of participants, providing insights into how camera and editing techniques influence gaze patterns associated with objectification. To further visualize these patterns, we employed Sequence Chart \citep{blascheck2014state}.  

\begin{figure}[t] 
    \centering
    \includegraphics[width=0.99\columnwidth]{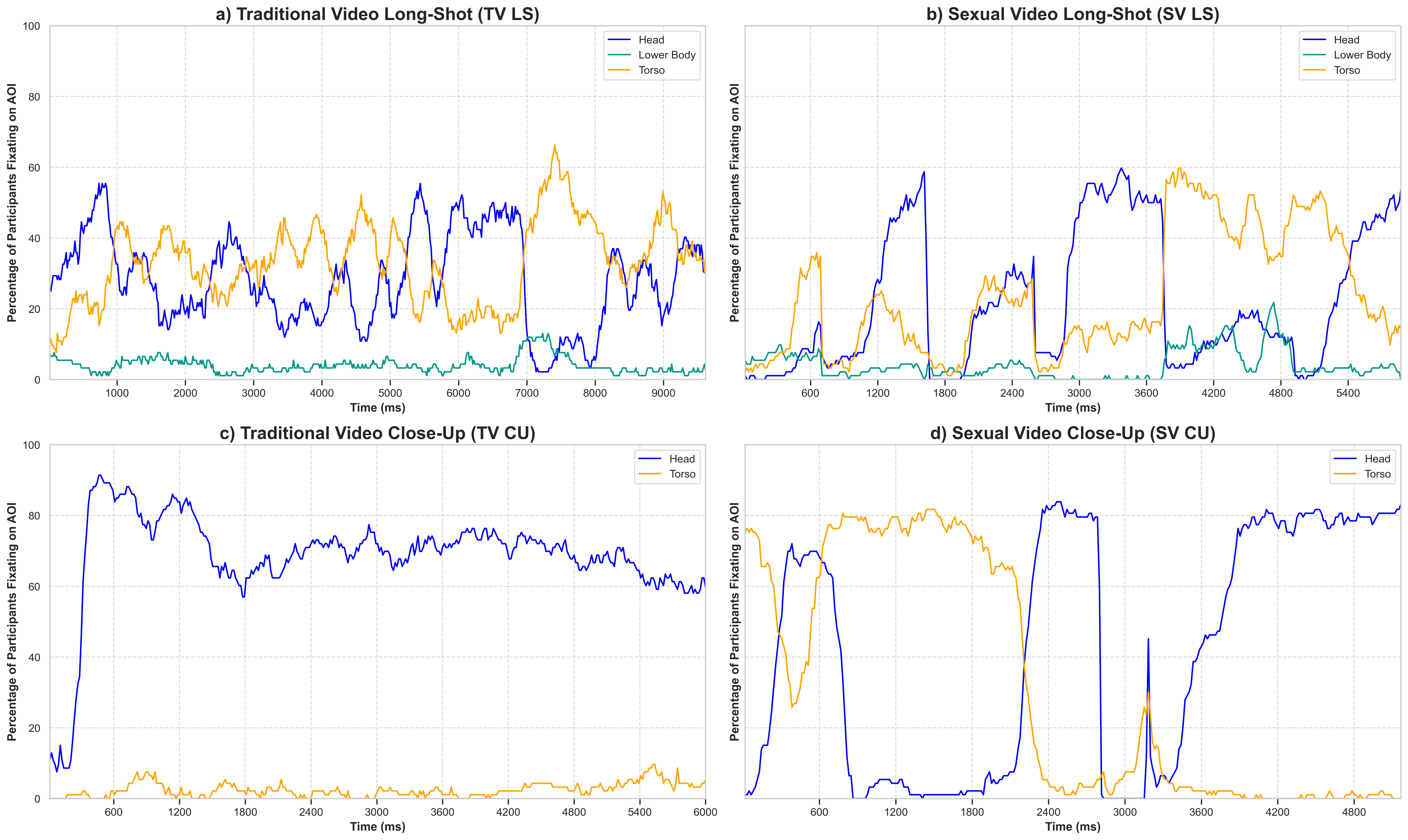} 
    \caption{Comparison of gaze patterns across AOIs (Head: blue, Torso: orange, Lower Body: green) over time between sexualized (SV) and non-sexualized (TV) videos for long-shots (LS) and close-ups (CU).}
    \label{fig:aoi_plot}
\end{figure}

\section{Results}

\subsection{Attentional Distribution across AOIs for scenes}
Figure 3 illustrates the temporal dynamics of viewer's attention to each AOI for in selected scenes. As shown in Figure 3c and 3d,  the average fixation duration on the torso was significantly higher in the SV CU (1729.19 ms) compared to the TV CU (153.49 ms), indicating a greater tendency towards objectification. The SV LS (Figure 3b) shows a substantial portion of viewers focusing on the torso and lower body AOIs for extended periods. SV CU also exhibited large-scale transitions, with 70-80\% of participants shifting their attention between the head and torso, indicating camera-driven objectification. In contrast, TV LS (Figure 3a) showed less severe shifts and a more even distribution of attention across AOIs. Moreover, TV CU scenes consistently engaged a similar majority on the head, without guided attention to the torso. 

\subsection{Gaze Behavior across scene conditions using Statistical Tests}

\subsubsection{Fixation Duration:} Wilcoxon tests (Table 1) on fixation duration reveal distinct attentional allocation patterns. For CU condition, participants fixated significantly longer on the head than on the torso in the TV condition (p $<$ 0.001), whereas SV scenes garnered more attention on torso (p $<$ 0.001). Comparing SV CU and TV CU, significant differences emerged for both head (p $<$ 0.001) and torso (p $<$ 0.001), indicating that sexualized framing redistribute attention. In LS condition, while some non-significant differences in head vs. torso fixation duration emerged (e.g., SV LS head vs. torso, p $>$ 0.05), yet pronounced disparities emerged when comparing sexualized vs. non-sexualized contexts for head (p $<$ 0.05) and torso AOIs (p $<$ 0.001). These results confirm that sexualized camera techniques shape attentional distributions, directing focus toward sexualized AOIs.

\subsubsection{Visit Count:}
Visit count analyses show similar trends. In CU scenarios, participants revisited the head AOI more frequently in TV than SV (p $<$ 0.001), reflecting a stable preference for non-sexualized targets when the scene is less suggestive. Conversely, in SV CU scenes, participants exhibited a greater tendency to revisit the torso (p $<$ 0.001). In LS conditions, participants under SV frequently revisited torso and hips/legs AOIs, diverging significantly from TV LS patterns (p $<$ 0.001). Between conditions, head and torso visit counts differ significantly in many cases, reinforcing the notion that sexualized camera angles and editing styles prompt more dynamic scanning strategies favoring sexualized AOIs. Overall, these visit count patterns align with the fixation duration results, providing converging evidence that sexualized framing induces systematic, measurable shifts in attentional behavior.

\begin{table}[h!]
\centering
\caption{Wilcoxon Test Results for Fixation Duration (FD) and Visit Count (VC) across AOIs (H: Head, T: Torso, LB: Lower Body) and Conditions. Values for \textbf{Fixation Duration (FD)} are \textcolor{blue}{\textbf{blue}} and \textbf{Visit Count (VC)} are \textcolor{teal}{\textbf{teal}}.}
\vspace{1em}
\small
\renewcommand{\arraystretch}{1.1}
\setlength{\tabcolsep}{2pt}
\begin{tabular}{llll}
\hline
\textbf{Condition} & \textbf{AOI} & \multicolumn{1}{c}{\textbf{z-score}} & \multicolumn{1}{c}{\textbf{p-value}} \\
\hline
SV CU & H vs. T & \textcolor{blue}{\hspace{5pt}689.41} \hspace{11pt} \textcolor{teal}{589.41} & \textcolor{blue}{\hspace{3pt}$<$0.001} \hspace{5pt} \textcolor{teal}{$<$0.001} \\
TV CU & H vs. T & \textcolor{blue}{\hspace{5pt}-1.09} \hspace{21pt} \textcolor{teal}{46.41} & \textcolor{blue}{\hspace{3pt}$<$0.001} \hspace{5pt} \textcolor{teal}{$<$0.001} \\
CU (SV vs. TV) & H & \textcolor{blue}{\hspace{5pt}185.91} \hspace{6pt} \textcolor{teal}{1752.91} & \textcolor{blue}{\hspace{3pt}$<$0.001} \hspace{7pt} \textcolor{teal}{0.0981} \\
CU (SV vs. TV) & T & \textcolor{blue}{\hspace{5pt}33.91} \hspace{15pt} \textcolor{teal}{321.91} & \textcolor{blue}{\hspace{3pt}$<$0.001} \hspace{5pt} \textcolor{teal}{$<$0.001} \\
SV LS & H vs. T & \textcolor{blue}{\hspace{5pt}1782.41} \hspace{1pt} \textcolor{teal}{1396.91} & \textcolor{blue}{\hspace{5pt}0.4677} \hspace{8pt} \textcolor{teal}{0.1167} \\
SV LS & T vs. LB & \textcolor{blue}{\hspace{5pt}96.41} \hspace{19pt} \textcolor{teal}{79.41} & \textcolor{blue}{\hspace{3pt}$<$0.001} \hspace{5pt} \textcolor{teal}{$<$0.001} \\
SV LS & H vs. LB & \textcolor{blue}{\hspace{5pt}354.91} \hspace{10pt} \textcolor{teal}{393.91} & \textcolor{blue}{\hspace{3pt}$<$0.001} \hspace{5pt} \textcolor{teal}{$<$0.001} \\
TV LS & H vs. T & \textcolor{blue}{\hspace{5pt}1929.91} \hspace{1pt} \textcolor{teal}{1756.41} & \textcolor{blue}{\hspace{5pt}0.4179} \hspace{8pt} \textcolor{teal}{0.7590} \\
TV LS & T vs. LB & \textcolor{blue}{\hspace{5pt}160.91} \hspace{14pt} \textcolor{teal}{86.41} & \textcolor{blue}{\hspace{3pt}$<$0.001} \hspace{5pt} \textcolor{teal}{$<$0.001} \\
TV LS & H vs. LB & \textcolor{blue}{\hspace{5pt}307.91} \hspace{10pt} \textcolor{teal}{172.91} & \textcolor{blue}{\hspace{3pt}$<$0.001} \hspace{5pt} \textcolor{teal}{$<$0.001} \\
LS (SV vs. TV) & H & \textcolor{blue}{\hspace{5pt}1301.91} \hspace{1pt} \textcolor{teal}{1150.91} & \textcolor{blue}{\hspace{3pt}$<$0.05} \hspace{9pt} \textcolor{teal}{$<$0.001} \\
LS (SV vs. TV) & T & \textcolor{blue}{\hspace{5pt}1145.91} \hspace{1pt} \textcolor{teal}{1856.91} & \textcolor{blue}{\hspace{3pt}$<$0.001} \hspace{7pt} \textcolor{teal}{0.3523} \\
\hline
\end{tabular}
\label{tab:metrics_tests}
\end{table}

\begin{figure*}[h] 
    \centering
    \includegraphics[width=0.96\textwidth]{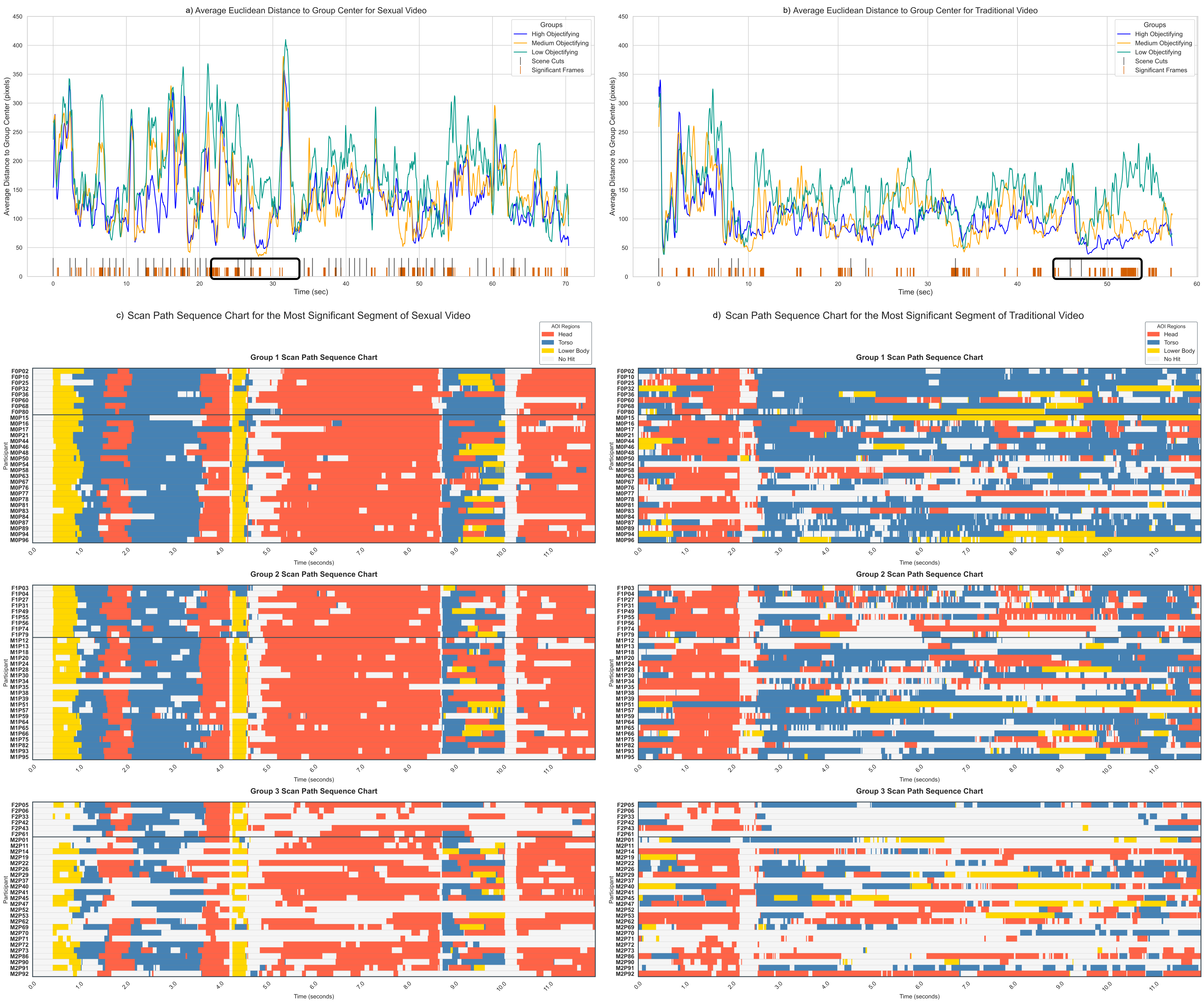} 
    \caption{\textbf{Gaze behavior across sexualized (SV) and traditional (TV) videos}. (4a \& 4b) Average Euclidean distance to group center over time for TV and SV. Lower values indicate greater gaze coherence within group. Red/grey vertical bars indicate significant differences/scene transitions, respectively. (4c \& 4d) Scan path sequence charts for high, medium, and low objectifying groups for the most significant segments of SV and TV. Solid same-color blocks indicate similar gaze patterns.}
    \label{fig:fig4and5}
\end{figure*}

\subsection{Significant Segment Selection for exploring Gaze Coherence}
Figure 4 (a \& b) presents the temporal evolution of group-level gaze coherence across subjects for both TV and SV conditions, comparing three participant groups (high, medium and low objectifying group). The y-axis represents gaze dispersion, with lower values indicating group members looking at similar regions onscreen, and higher values signaling more diverse, individualized viewing patterns.
In the TV condition (Figure 4b), gaze coherence fluctuates widely. Although brief intervals of lower dispersion appear, the general trend shows each group’s gaze evolving more independently over time. Scene transitions occasionally trigger transient alignment in participant’s gaze before they revert to their idiosyncratic scanning behaviors. This suggests that, without overt sexualization cues, the camera’s framing exerts weaker control over where viewers look, leading to more varied attentional strategies.
By contrast, the SV condition (Figure 4a) reveals a more pronounced influence of objectifying camera techniques. Over several intervals, all three groups—despite differing baseline objectification tendencies—exhibit synchronized dips in dispersion, clustering their gaze on similar AOIs. Notably, even the group initially least inclined toward sexualized AOIs (Group 3, green line) gradually converges toward these focal points after scene transitions. These temporal “pulls” highlight how camera framing and editing rhythms can override initial individual differences, guiding attention into more uniform, objectifying gaze distributions. These findings not only confirm that dynamic camera cues can modulate attentional focus but also suggest that scene transitions act as cognitive “reset points,” allowing camera-induced objectification effects to repeatedly emerge \citep{gervais2013my, Flynn2016ObjectificationIP, Valuch2014TheEO}.

\subsection{Gaze Synchronization in significant segments}
For each condition, we initially identified multiple significant video segments using our data-driven segmentation technique. Across TV condition, we found 3 non-overlapping significant segments (mean segment duration $\approx$ 12–14 seconds), while the SV condition yielded 7 such non-overlapping segments. After manual inspection, we selected two representative 12-second intervals for each condition—approximately 45–57 seconds into the TV footage and 22–34 seconds into the SV footage—for detailed analysis. We then encoded participants’ scanpaths as AOI sequences (A=Head, B=Torso, C=LowerBody, X=No AOI Fixation) and created the ScanGraphs \citep{Dolezalov2016ScanGraphAN}.
The ScanGraph\footnote{ScanGraph for SV Segment \textbf{\href{https://eyetracking.upol.cz/scangraph/?source=1112790962679768c6c3c256.42883717}{https://tinyurl.com/SV-sps-22}} and TV Segment \textbf{\href{https://eyetracking.upol.cz/scangraph/?source=1035305008679768b5134010.27826688}{https://tinyurl.com/TV-sps-45}} (Load time $\approx$ 15 min)} for the SV segment revealed a dense cluster of 52 participants, composed of overlapping subgroups of up to 16 participants, all converging on similar objectifying gaze patterns. This high edge count (483 edges, $\sim$12\%) signifies a strong, unified attentional bias towards sexualized AOIs. Moreover, at critical timestamps (e.g., $\sim$9s), participants uniformly shifted focus to torso and hips/legs AOIs, even when head AOIs were equally available. White spaces in scanpaths—possibly due to camera cuts or AOIs moving off-screen—often preceded fixation on these sexualized regions, underscoring the camera’s role in continuously re-engaging viewers with these focal points (as shown in Figure 4 c \& d). 
In contrast, ScanGraph for TV segment showed fewer edges (80 edges, $\sim$2\%) and produced only small separate clusters of about 14 and 7 participants each, reflecting more varied, individualized viewing strategies without a dominant pattern. This fragmentation suggests that participants explored available AOIs more freely—often lingering on the head or scanning the scene’s dynamic gestures—leading to heterogeneous gaze sequences (as shown in Figure 5).
Overall, these results indicate that sexualized camera framing can override baseline group differences, herding a majority of viewers into uniform, objectifying gaze patterns, whereas more traditional framing affords greater freedom and diversity in attentional allocation.

\section{Discussion}



The sexualized camera framing under SV condition directs uniform, objectifying gaze patterns, whereas the TV scenario allows for more diverse and individualized viewing strategies, as illustrated in Figure 3. Enhanced fixation duration and visit counts directed towards sexualized AOIs in the SV condition answer \textbf{RQ1}, extending prior research \citep{FredricksonRoberts1997, Bartky1990, Karsay2018AdoptObj} by demonstrating that sexualized portrayals not only encourage objectification but also align viewers’ gaze with camera cues, reinforcing and normalizing fragmented perceptions of women. Segmentation analysis identified temporal intervals where objectifying attention intensified, often during scene cuts or CU shots emphasizing sexualized body parts, suggesting that brief editing choices can amplify focus on objectifying elements \citep{Canini2011AffectiveAO}. CU shots more effectively narrowed viewer's attention to sexualized AOIs than LS, answering \textbf{RQ2} by demonstrating how tighter framing limits visual freedom and directs gaze to specific body regions. These findings highlight how visual composition and editing techniques shape viewer engagement and reinforce objectifying perceptions.

The higher gaze synchronization across participants observed in the SV condition, as shown in Figure 4 (c \& d), where they focused on similar AOIs at nearly the same time, compared to the more varied gaze patterns in TV, validates \textbf{H1}. This collective shift indicates that objectifying camera framing not only impacts individual cognition but also fosters shared viewing dynamics, where personal biases may get overridden by uniform scanning behavior. The ScanGraph and sequence chart analyses provide complementary insights into this synchrony. While the ScanGraph visualizes clustering patterns and their density, revealing the degree of synchronization, sequence charts offer a detailed view of individual scanpaths, showing the temporal shifts in attention to specific AOIs. SV segments, as seen in the sequence charts, reveal dense clusters in the ScanGraph, with participants from all objectification groups (low, medium, high) focusing on sexualized AOIs at similar times. This finding is significant as it shows that even those relatively less predisposed to objectification are still influenced by sexualized framing. Cultural and cinematic factors can override baseline tendencies, aligning viewer attention and reinforcing objectification, extending previous research \citep{loughnan2010objectification, Ward03052016} that media environments reshape attentional biases, regardless of individual dispositions.


These results offer a compelling narrative that sexualized camera techniques not only direct but also synchronize attention, creating shared objectifying perceptions that override individual variablility. By integrating eye-tracking, dynamic segmentation, and culturally specific media forms, this study opens new avenues for examining the subtleties of media-induced objectification. Our findings reinforce the cognitive and cultural aspects of objectification theory and stress the importance of studying dynamic, visually rich media. This understanding can guide future research on how subtle cinematic techniques shape cognition, adversely affect perceptions of women, and perpetuate objectifying norms, urging socially responsible cinematography.

\section{Limitations and Future Research}

Our study's scope is inherently bound by a few factors, providing fertile ground for future research. While our focus on a South-Asian cultural context and specific dance forms offers a critical vantage point for examining non-Western media \citep{bhandari2018commodification, loughnan2010objectification}, future investigations could explore the generalizability of these findings across diverse cultural contexts and cinematic styles. Expanding the range of stimuli by analyzing multiple videos of different dance forms and incorporating more fine-grained analyses of multimodal cues, such as hand/leg gestures and auditory elements, would further enhance our understanding of the complex interplay between cinematic techniques and viewer attention. The generalizability of our study is limited by our sample of engineering students from a single Indian university, warranting future research with more diverse populations. Additionally, incorporating physiological measures, such as sexual arousal, could provide richer insights into the cognitive and affective mechanisms underlying objectification. Holistically, our work lays a foundation for more inclusive, nuanced investigations that bridge cognitive science, technology, and cultural analysis.

\printbibliography

\end{document}